\documentclass[10pt,aps,pra,twocolumn,superscriptaddress,noeprint]{revtex4-2}
\usepackage{graphicx}
\usepackage{url}
\usepackage{amsmath,amsthm,amssymb,physics,color,enumerate,tabularx,makecell,mathrsfs,float,bbm,stackrel,mathtools,multirow,bm}
\usepackage{natbib}
\usepackage{comment}

\usepackage[dvipsnames]{xcolor} % colours
\usepackage[colorlinks=true,%
bookmarks=false,%
linkcolor=blue,%
urlcolor=blue,%
citecolor=blue,%
breaklinks]{hyperref}
\usepackage{orcidlink}

\newcommand{\free}{\mathcal{F}_\beta}
\newcommand{\work}{\mathcal{W}^\beta}
\newcommand{\erg}{\mathcal{E}}
\newcommand{\daem}{\overline{\mathcal{E}}_{\{\Pi_a^A\}}}
\newcommand{\enhanwork}{\overline{\mathcal{W}}^\beta_{\{\Pi_a^A\}}}

\begin{document}
\title{Energetics in daemonic work extraction protocols via non-ideal QND-energy measurement}

\author{Daniele Morrone\, \orcidlink{0000-0001-9934-9971}}
\email{daniele.morrone@upol.cz}
\affiliation{Università degli Studi di Palermo, Dipartimento di Fisica e Chimica - Emilio Segrè, via Archirafi 36, I-90123 Palermo, Italy}
\author{Francesco Albarelli\,\orcidlink{0000-0001-5775-168X}}
\email{francesco.albarelli@gmail.com}
\affiliation{Università di Parma, Dipartimento di Scienze Matematiche, Fisiche e Informatiche, I-43124 Parma, Italy}
\affiliation{INFN—Sezione di Milano-Bicocca, Gruppo Collegato di Parma, I-43124 Parma, Italy}
\author{Vittorio Giovannetti}
\affiliation{NEST-CNR Scuola Normale Superiore, Piazza dei Cavalieri 7, I-56126 Pisa, Italy}
\author{Mauro Paternostro}
\affiliation{Università degli Studi di Palermo, Dipartimento di Fisica e Chimica - Emilio Segrè, via Archirafi 36, I-90123 Palermo, Italy}
\affiliation{Centre for Quantum Materials and Technologies, School of Mathematics and Physics, Queen's University Belfast, BT7 1NN, United Kingdom}
\author{Marco G. Genoni\,\orcidlink{0000-0001-7270-4742}}
\email{marco.genoni@unimi.it}
\affiliation{Dipartimento di Fisica {\it "Aldo Pontremoli"}, Universit\`a degli Studi di Milano, I-20133 Milano, Italia}

\begin{abstract}
We address the problem of extracting work from a quantum system assisted by a quantum non-demolition (QND) energy measurement. When a perfect QND measurement can be performed and an auxiliary zero-temperature bath is available, the full energy of the quantum state can in principle be extracted even without any prior information on the input  state. Owing to the presence of a zero-temperature bath, this is achieved at no energetic cost for the measurement process itself.
On the contrary, here we consider what happens when the same protocol is implemented in non-ideal scenarios, specifically when the auxiliary bath has a finite temperature. In this case, not only is it impossible to extract the entire energy from the system, but the measurement strategy also acquires a non-zero energetic cost, accounting for both the interaction between system and  measurement apparatus, and the corresponding Landauer erasure cost. 

We quantitatively assess the performance of these work-extraction protocols, both in absolute terms and through the so-called {\it daemonic net gain}, which explicitly includes the energetic cost of the measurement.
We rigorously prove that, when access to a thermal bath is allowed in the extraction protocol, the daemonic net gain is always non-positive for any temperature of the auxiliary bath. Conversely, when only unitary operations are considered, the daemonic net gain can attain positive values.
We further discuss these different figures of merit by analyzing a paradigmatic example for a single qubit system.

\end{abstract}
\maketitle
\section{Introduction}
\label{s:intro}
The extraction of work from quantum systems is one of the central problems of quantum thermodynamics~\cite{binderThermodynamicsQuantumRegime2018, deffnerquantumthermodynamics2019,vinjanampathyQuantumThermodynamics2016}.
For finite-dimensional systems, the maximum work extractable under unitary control is quantified by ergotropy~\cite{allahverdyanMaximalWorkExtraction2004a}, while access to a thermal bath leads to work bounds formulated in terms of non-equilibrium free-energy differences and thermal operations~\cite{Skrzypczyk2014,biswasExtractionErgotropyFree2022,faistGibbspreservingMapsOutperform2015,lostaglioElementaryThermalOperations2018}.
These notions are particularly relevant for finite-size quantum devices, including quantum batteries~\cite{alickiEntanglementBoostExtractable2013b,campaioliColloquiumQuantumBatteries2024,FerraroNatRevPhys2026}, where the extractable work depends not only on the energetic content of the state, but also on the operations and information available to the agent performing the extraction.
Indeed, optimal protocols generally require knowledge of the input state.
When such knowledge is partial or absent, the operationally accessible work can be reduced, and recent studies have addressed ergotropy extraction, estimation, and guarantees under incomplete state information~\cite{safranekWorkExtractionUnknown2023a,Joshi2025,Canzio2025a,Chakraborty2025a,Biswas2026,Pagliaro2026}.
In the asymptotic many-copy regime universal protocols may recover the optimal free-energy rate without prior state knowledge~\cite{Watanabe2026}, while in finite-copy settings learning the state and extracting work remain competing tasks~\cite{Lumbreras2025a}.

Measurements are useful in this context because they can provide information about the state, but their thermodynamic role is more fundamental.
Even when the initial state is perfectly known, allowing measurements and outcome-dependent feedback enlarges the class of admissible extraction protocols.
A single deterministic unitary can only extract the ergotropy of the input state, while a measurement can decompose the same state into conditional states from which more work can be extracted on average.
This mechanism is at the basis of measurement-assisted work extraction and of the notion of daemonic ergotropy~\cite{francicaDaemonicErgotropyEnhanced2017,bernardsDaemonicErgotropyGeneralised2019a}, and has been explored in different finite-dimensional and quantum-battery settings~\cite{manzanoOptimalWorkExtraction2018a,morroneDaemonicErgotropyContinuously2023,elyasiExperimentalSimulationDaemonic2024,cenedese2026}.
In the specific case considered in this work, the relevant measurement is a quantum non-demolition (QND) measurement of the system energy~\cite{ralphQuantumNondemolitionMeasurements2006,nakajimaQuantumNondemolitionMeasurement2019}.
As we will rigorously show in the manuscript, in the ideal limit, such a measurement can reveal the occupied energy eigenspace without directly destroying the system.
Choosing the ground-state energy as the zero of energy, an ideal QND energy measurement is sufficient, in principle, to extract the full mean energy of the initial state: each outcome identifies a conditional energy eigenstate, and a feedback operation can then map that state to the ground state while extracting the corresponding energy.
Averaging over outcomes gives the initial average energy, independently of any prior knowledge of the input state.

Such ideal conclusion hides an important thermodynamic subtlety.
The measurement apparatus, or equivalently the auxiliary system used to perform the QND measurement, must be reset to close the protocol.
The energetic cost of measurement and erasure is a central ingredient in the thermodynamics of information and Maxwell-daemon-like settings~\cite{sagawaSecondLawThermodynamics2008,abdelkhalekFundamentalEnergyCost2018b,jacobsQuantumMeasurementFirst2012b,deffnerQuantumWorkThermodynamic2016b,CottetPNAS2017,ferrazWeakContinuousMeasurements2025,latuneThermodynamicallyConsistentApproach2025,minagawaUniversalValiditySecond2024}.
Within the usual Landauer accounting, however, an ideal QND measurement initialized and reset by a zero-temperature bath can be assigned zero energetic reset cost: the bath prepares the auxiliary system in a pure ground state, and the factor proportional to the bath temperature vanishes.
This is a useful idealization, and it is often an excellent effective description of experiments in which the relevant quantum system is coupled to a very cold environment.
Nevertheless, the zero-temperature bath is itself a thermodynamic resource.
At the level of an effective master equation the cooling apparatus is typically not modeled explicitly, but in many physical implementations maintaining the environment at very low temperature carries a hidden macroscopic energetic cost.

\begin{figure}
	\centering
		\includegraphics[width=0.48\textwidth]{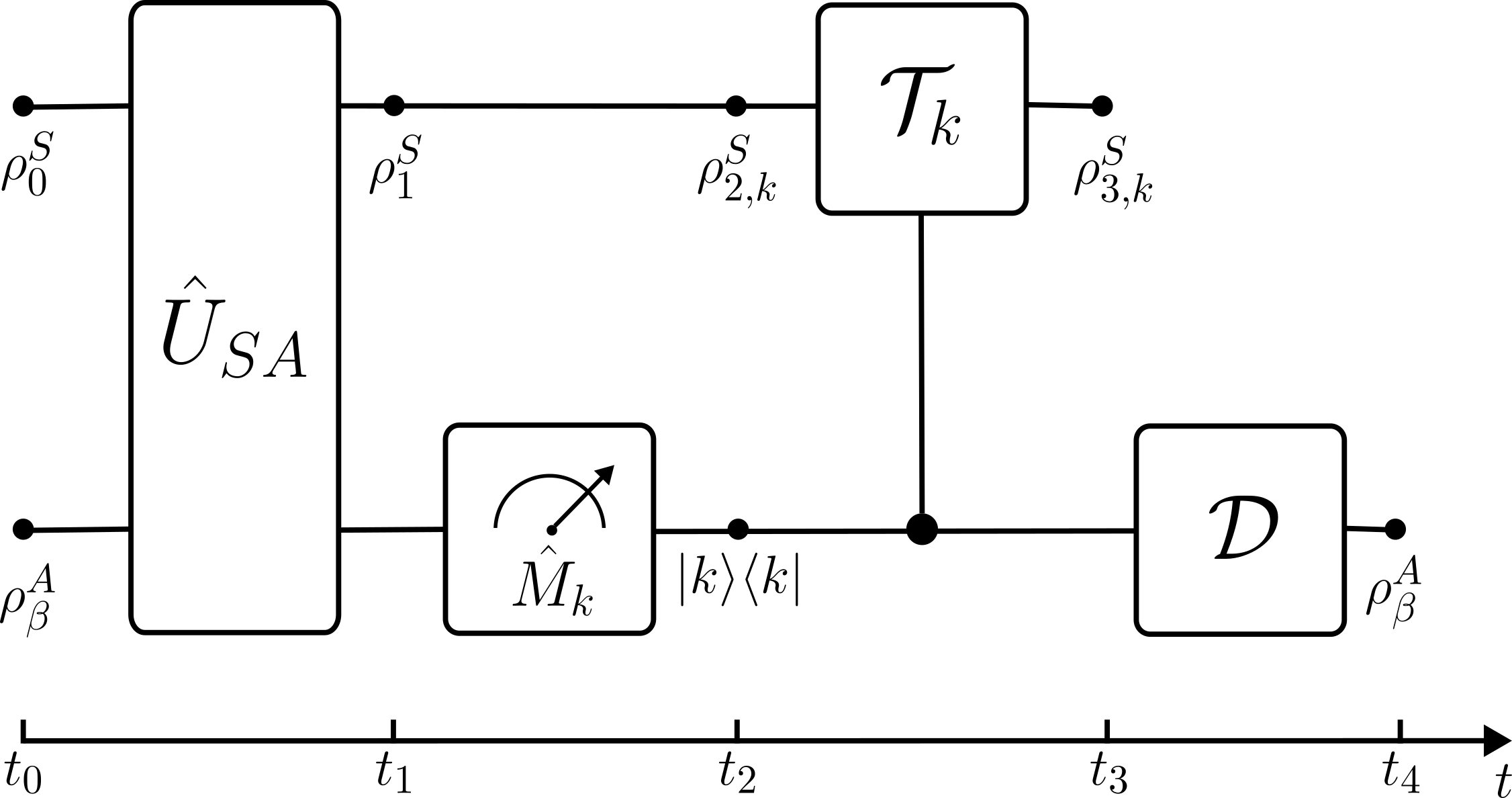}  
	\caption{A schematics for the QND energy measurement protocol we investigate. The main system \(S\) is made to interact with an auxiliary system to create correlation between them. The auxiliary system \(A\) is then projected through a set of PVM operators, allowing indirect measurement of the system \(S\), where the amount of information gained is inversely proportional to
    the temperature of the initial state of \(A\): for the system's state initially in the zero temperature thermal state (ground state), the maximum amount of information is obtained, while for the infinite temperature state no information at all is obtained.
    Based on the measurement outcome, an extraction operation is performed on $S$, which could be either a purely unitary operation or one potentially aided by access to a thermal bath. Finally, the auxiliary system is reset to the thermal state. For step 3\&4, a thermal bath may be involved to realize the operation.}
	\label{f:qnd_model}
\end{figure}

This observation motivates the question addressed here: what changes when the bath available for resetting the measurement auxiliary system is not at zero temperature?
We study this question in a minimal model of noisy QND energy measurements.
As sketched in Fig.~\ref{f:qnd_model}, the system \(S\) is coupled to an auxiliary system \(A\), the auxiliary system is measured projectively, and the outcome is used to choose the subsequent work-extraction operation.
The non-ideality is introduced by preparing the auxiliary system in a thermal state rather than in its ground state.
Thus the same bath temperature controls both the noise in the QND measurement and the thermodynamic cost of resetting the auxiliary system.
We deliberately assume that there is a single bath at a single temperature throughout the protocol: the finite-temperature bath used to reset the measurement apparatus is the same thermal resource that may be exploited in the work extraction.
In this way no additional work can be gained by implicitly combining reservoirs at different temperatures or by building an auxiliary heat engine.

We analyze this finite-temperature QND-assisted extraction protocol for generic finite-dimensional systems and then specialize the results to a qubit system measured through a qubit auxiliary system.
Our main comparison is between two operational settings.
In the first, the feedback extraction after the noisy QND measurement is restricted to unitary operations on the system, leading to a daemonic-ergotropy figure of merit.
In the second, in the feedback extraction one has access to a thermal bath with the same temperature as the reset bath, leading to a measurement-assisted free-energy work figure of merit.
In both cases we compare the gross measurement-assisted extracted work with the corresponding no-measurement benchmark, and we also include the energetic cost of the QND measurement to assess the net daemonic gain.

The paper is organized as follows.
In Section~\ref{s:theory} we review the notions needed for the analysis: unitary and thermal work extraction, measurement-enhanced work, ideal QND energy measurements, and the measurement-cost framework used throughout the paper.
In Section~\ref{s:res1} we introduce the finite-temperature QND protocol and derive general results for thermal and unitary extraction.
In Section~\ref{s:res2} we specialize the protocol to a qubit-qubit implementation and discuss the resulting figures of merit explicitly.
Finally, in Section~\ref{s:conclusion} we summarize the results and outline possible extensions.

\section{Theoretical framework}
\label{s:theory}
In this Section we review theoretical concepts used throughout the paper. In order to fix the ideas, as we are mainly interested in work extraction, in what follows we adopt the convention according to which work done by the system is considered as positive.

\subsection{Quantum work definitions}
We introduce two different work definitions based on the type of operations allowed to perform the extractions: unitary operations acting on the system only and operations where access to a thermal bath is allowed.

Work extraction with unitary operation is known as ergotropy, as introduced in Ref.~\cite{allahverdyanMaximalWorkExtraction2004a}. For a finite-size quantum system with Hamiltonian $\hat{H}_S$ and initially in a generic state $\rho_0$, the amount of work extracted through a unitary process $\hat{U}$ is
\begin{align}
	\mathcal{W}_U=E(\rho_0)-E(\hat{U}\rho_0 \hat{U}^\dagger) \label{eq:unitary_work}
\end{align}
where $E(\rho)=\Tr\left[ \hat{H}_S \rho \right]$. To quantify how much work we can actually extract from the state $\rho_0$, formally, we need to solve the following optimization problem
\begin{align}
	\mathcal{E}(\rho_0)=\max_{\hat{U}} \mathcal{W}_U, \label{eq:ergotropy}
\end{align} 
where the quantity $\mathcal{E}$ is known as ergotropy. However, the optimal unitary sends the initial state to its passive state \(\tau_\rho\), which is easily identifiable. Given the spectral decomposition of the state $\rho_0$
\begin{align}
	\rho_0=\sum_j r_j \ketbra{r_j}, \quad \quad r_1\geq r_2 \geq ...
	\label{eq:rho_0}
\end{align}
and the spectral decomposition of the Hamiltonian $\hat{H}_S$
\begin{align}
	\hat{H}_S= \sum_k \varepsilon_k \ketbra{\varepsilon_k}, \quad \quad \varepsilon_1 \leq \varepsilon_2 \leq ...
	\label{eq:h_s}
\end{align}
where, without loss of generality, we set the lowest eigenvalue equal to zero, the corresponding passive state reads
\begin{align}
	\tau_{\rho_0}=\sum_j r_j \ketbra{\varepsilon_j},
\end{align}
which can be obtain by applying the unitary operation
\begin{align}
	\hat{U}=\sum_j \ketbra{\varepsilon_j}{r_j}. \label{eq:unitary_extr}
\end{align}
This allows us to directly compute Eq.~(\ref{eq:ergotropy}) as
\begin{align}
	\mathcal{E}(\rho_0)=\sum_{j,k} r_j \varepsilon_k\left( \abs{\braket{r_j}{\varepsilon_k}}^2 -\delta_{j,k} \right).
	\label{eq:ergotropy_explicit}
\end{align} 
It can also be useful to write the ergotropy in terms of the more familiar non equilibrium free energy $\mathcal{F}_\beta$
\begin{align}
	\mathcal{F}_\beta(\rho)=E(\rho)- \frac{1}{\beta} S(\rho),
\end{align}
where $S(\rho)=-\Tr \left[ \rho \ln (\rho) \right]$ is the Von Neumann entropy. Because unitary operation preserve the entropy of the state, we can write
\begin{align}
	\mathcal{E}(\rho_0)&=E(\rho_0)-E(\tau_{\rho_0}) \nonumber \\
	&=E(\rho_0)+\frac{1}{\beta} S(\rho_0)-\frac{1}{\beta} S(\rho_0)-E(\tau_{\rho_0})\nonumber \\
	&=\mathcal{F}_\beta(\rho_0)-\mathcal{F}_\beta(\tau_{\rho_0}), \label{eq:erg_as_freenergy}
\end{align}
where we used $S(\rho_0)=S(\tau_{\rho_0})$. In this equation, $\beta$ is a free parameter with no particular physical interpretation. 

We now consider the second setting, in which work extraction can also
exploit a thermal bath \(B\) at inverse temperature \(\beta\), described by the
Gibbs state \(\rho^B_\beta = e^{-\beta \hat{H}_B}/\Tr[e^{-\beta \hat{H}_B}]\).
Several operational protocols are conceivable here, depending on the set of
allowed operations~\cite{biswasExtractionErgotropyFree2022}.
Whichever protocol is chosen, as long as the system is coupled to a single bath
at inverse temperature \(\beta\) and its Hamiltonian is cyclically restored, the
average extracted work obeys the second law
\begin{align}
	&\work(\rho_0^S) \leq -\Delta \free,
	\label{eq:thermalwork}
\end{align}
where we defined
\begin{align}
	\Delta \free = \free(\rho_\beta^S)- \free(\rho^S_0).
\end{align}
We refer to the work extracted in this setting as ``isothermal work'', and we denote it by $\work$.
The bound~\eqref{eq:thermalwork} is attained, in the limit of an infinitely large bath and of quasi-static driving, by an operationally reversible isothermal protocol: a state-dependent sudden quench
\(\hat{H}_S \to -\beta^{-1}\ln \rho_0^S\), followed by a quasi-static isothermal
driving back to \(\hat{H}_S\) while the system remains in contact with the bath, the average extracted work being identified with \(W = -\int \Tr[\rho \, \dot{\hat{H}}]\, dt\)~\cite{manzanoOptimalWorkExtraction2018a}.
Throughout this work we refer to this ideal setting as \emph{reversible isothermal work extraction}, and we accordingly take
\begin{align}
    \work(\rho^S) = \free(\rho^S) - \free(\rho^S_\beta) \,.
    \label{eq:isothermalwork_def}
\end{align}

The general bound~\eqref{eq:thermalwork} 
can be obtained by noticing that all these protocols form a subset of the unitary operations acting on the composite system  \(S+B\), so that the extracted work cannot exceed the ergotropy of the joint system-bath state~\cite{biswasExtractionErgotropyFree2022}:
\begin{align}
	\mathcal{E}(\rho^S_0\otimes\rho^B_\beta)&=
    \free(\rho_0^S\otimes\rho^B_\beta)-\free (\tau_{\rho^S_0\otimes\rho^B_\beta})\nonumber\\
	&=
    \free(\rho_0^S\otimes\rho^B_\beta)-\free(\rho_\beta^S\otimes\rho^B_\beta) + \nonumber \\
	&+\free(\rho_\beta^S\otimes\rho^B_\beta)-\free (\tau_{\rho^S_0\otimes\rho^B_\beta}) \nonumber \\
	&=\frac{1}{\beta}S(\rho^S_0||\rho^S_\beta)- \frac{1}{\beta}S(\tau_{\rho^S_0\otimes \rho_\beta^B}||\rho^S_\beta\otimes \rho_\beta^B) \nonumber \\
	&\leq \frac{1}{\beta} S(\rho^S_0||\rho^S_\beta)=\free(\rho^S_0)-\free(\rho_\beta^S), \label{eq:work}
\end{align}
where
\begin{align}
	S(\rho||\tau)=\Tr \left[ \rho\left(\ln \rho - \ln \tau\right) \right]\,,
\end{align}
is the quantum relative entropy.
The last inequality returns the free-energy bound, which is tight only in the limit of an infinitely large bath, the setting we consider in this work.

\subsection{Measurement-enhanced quantum work}
Bounds on the work enhancement achievable exploiting measurements have been evaluated for both unitary and bath-assisted operations. To review these bounds we assume the following setting: In a bipartite system \(SA\) with initial state correlations, measurement can be performed on \(A\), following which an outcome-dependent thermal/unitary operation can be applied on \(S\). If we perform a measurement on the auxiliary system \(A\), described by a POVM $\{\hat{\Pi}_a^A\}$, the state of the system \(S\) unravels into a convex combination of conditional states as 
\begin{align}
	\rho^{S}= \sum_a p_a \rho^{S|a}\,,
\end{align}
where $p_a = \Tr_{SA}[\rho^{SA} (\hat{\mathbbm{1}}^S \otimes \hat{\Pi}_a^A)]$ is the probability of obtaining the measurement outcome $a$ and $\rho^{S|a} = \Tr_A[\rho^{SA} (\hat{\mathbbm{1}}^S \otimes \hat{\Pi}_a^A)]/p_a$ is the corresponding conditional state. The ergotropy of the conditional state \(\rho^{S|a}\) is expressed as
\begin{align}
	\mathcal{E}(\rho^{S|a})=E(\rho^{S|a})- \Tr \left[\hat{H}_s \hat{U}_a \rho^{S|a} \hat{U}_a^\dagger\right],
\end{align}
where \(\hat{U}_a\) is the unitary operation that optimizes work extraction for the measurement outcome \(a\). The daemonic ergotropy, defined in~\cite{francicaDaemonicErgotropyEnhanced2017} as the average over all possible measurement outcomes, is given by:
\begin{align}
	\overline{\mathcal{E}}_{\{ \Pi_a^A \}}&= \sum_a p_a \mathcal{E} (\rho^{S|a}) \label{eq:daem_mean} \\
	&= E(\rho^S) -  \sum_a p_a  \Tr \left[ \hat{H}_s \hat{U}_a \rho^{S|a}\hat{U}_a^\dagger \right],
\end{align}
where we used the linearity of the trace to simplify the first term. It is useful to define the daemonic gain as
\begin{align}
	\Delta\overline{\mathcal{E}}_{\{ \Pi_a^A \}} =  	\overline{\mathcal{E}}_{\{ \Pi_a^A \}} - \mathcal{E}(\rho^S) \geq 0 \,,
\end{align}
which represents the ergotropy gain due to the measurement. The positivity follows directly from the convexity of the ergotropy. The equality is achieved in the case when the initial state $\rho^{SA}$ is factorized
\begin{align}
	\rho^{SA}=\rho^S\otimes \rho^A,
\end{align}
or for separable initial states $\rho^{SA}$ that lead to conditional states $\rho^{S|a}$ characterized by the same ($a$-independent) optimal extraction unitary $\hat{U}$. For any other case the daemonic ergotropy is always an improvement to the ergotropy. 

For bath-assisted operations, a similar approach is presented in~\cite{manzanoOptimalWorkExtraction2018a}. If the maximum extractable work from the subsystem \(S\) is given by $\work(\rho^S)$, and we quantify the change in free energy due to a measurement with outcome \(a\) as
\begin{align}
	\Delta \mathcal{F}_\beta^{S|a}=\mathcal{F}_\beta (\rho^{S|a}) - \mathcal{F}_\beta (\rho^{S}),
\end{align}
then the total extractable work, conditioned on outcome \(a\), is given by
\begin{align}
	\mathcal{W}^\beta(\rho^{S|a})=\mathcal{W}^{\beta}{(\rho^S)}+\Delta \mathcal{F}_\beta^{S|a}.
\end{align}
The total work extracted, averaged over all outcomes, is:
\begin{align}
	\overline{\mathcal{W}}^\beta_{\{\hat{\Pi}_a^A\}}= \sum_a p_a \mathcal{W}^\beta(\rho^{S|a}) = \sum_ a p_a \Delta \mathcal{F}_\beta^{S|a} + \mathcal{W}^{\beta}{(\rho^S)}.
\end{align}
To take advantage of the enhancement, the protocol must be conditioned by the outcome of the measurement.

As with the previous result concerning the daemonic gain, the convexity of the free energy in this case also guarantees that
\begin{align}
	\Delta \overline{\mathcal{W}}^\beta_{\{\hat{\Pi}_a^A\}} &= \overline{\mathcal{W}}^\beta_{\{\hat{\Pi}_a^A\}} -  \mathcal{W}^{\beta}{(\rho^S)} \geq 0,
\end{align}
meaning that the gain due to measurement is always positive; as in the previous case, the lower bound is achieved only with initial states $\rho^{SA}$ that are factorized or separable and leading to conditional states characterized by the same extraction protocol.

An important result can be obtained if we explicitly evaluate the term $\sum_a p_a \Delta \mathcal{F}_\beta^{S|a}$. In fact, we get
\begin{align}
	\Delta \overline{\mathcal{W}}^\beta_{\{\hat{\Pi}_a^A\}} &
	= \sum_ a p_a \Delta \mathcal{F}_\beta^{S|a} \nonumber \\
	&=  \beta^{-1} \left( S(\rho^{S}) - \sum_a p_a S(\rho^{S|a}) \right)  \nonumber \\
	&= \beta^{-1} \mathcal{J}(\rho^{SA}) \,,\label{eq:work_gain}
\end{align}
where we applied the definition of $\mathcal{F}_\beta$ and used the linearity of the trace to simplify the term proportional to the energy. In the last line we substituted the quantity \( \mathcal{J}(\rho^{SA}) \), which is known as the quantum-classical mutual information and is defined as 
\begin{align}
	\mathcal{J}(\rho^{SA})=S(\rho^{S}) - \sum_a p_a S(\rho^{S|a})\,, \label{eq:qc_mut}
\end{align}
and thus corresponding to the Holevo information $\chi$ for the ensemble $\{p_a, \rho^{S|a}\}$.

\subsection{The ideal QND-energy measurement protocol}
Here, we simply characterize the ideal QND-energy measurement protocol as it will be applied later under non-ideal conditions. The measurement protocol involves two steps applied to a bipartite composite system \(SA\), where \(S\) is the system being measured. 

Crucially, \(S\) is not directly projected or altered in a way that destroys it, as happens in certain destructive measurements. For instance, photon detection requires its absorption by the detector, effectively destroying the photon. In contrast, in this QND scheme, the other component of the bipartite system, the auxiliary system \(A\), is projected instead of \(S\). 

In the ideal case, we assume that \(S\) is initially in an unknown state, while \(A\) starts in the ground state of its Hamiltonian \(\hat{H}_0^A\). We point out that for a regular QND measurement scheme, it is sufficient for the system \(A\) to be in a reference pure state, however we here choose to demonstrate the measurement scheme so that it is consistent with later assumptions. 
The initial state of the composite system is thus given by
\begin{align}
	\rho_0^{SA}=\rho_0^S \otimes \ketbra{0^A},
\end{align} 
where
\begin{align}
	\rho_0^S= \sum_{n,m} {\rho}_{n,m} 
    \ketbra{n}{m}\,,
\end{align}
represents a generic state of \(S\) expressed in the eigenbasis of the system's Hamiltonian \(\hat{H}_0^S = \sum_n \varepsilon_n \ketbra{n}\).
For simplicity, throughout this work we assume that the Hamiltonian $\hat{H}_0^S$ is non-degenerate.

The first step of the measurement process establishes a correlation between \(S\) and \(A\) by applying a unitary operation on the composite system. The unitary operator chosen is
\begin{align}
	\hat{U}_{SA}=e^{-i \hat{H}_0^S \otimes \hat{A}}, \label{eq:ch6_unitary}
\end{align}
with \(\hat{A}\) defined such that it satisfies:
\begin{align}
	\hat{A}: |\psi_n\rangle= e^{-i \varepsilon_n \hat{A}}|0\rangle, \label{eq:ch6_unitary_cond}
\end{align}
where the states \(\ket{\psi_n}\) form a orthogonal basis on the Hilbert space of \(A\). 

Under this unitary evolution, the state of the system becomes
\begin{align}
\rho_1^{SA} = \hat{U}_{SA} (\rho^S_0 \otimes \ketbra{0^A}) \hat{U}_{SA}^\dagger,
\end{align}
which expands as
\begin{align}
\rho_1^{SA} = \sum_{n,m} \rho_{n,m} 
\left( \ketbra{n^S}{m^S} \otimes \ketbra{\psi_n^A}{\psi_m^A} \right).
\end{align}
Noticeably, states of this form are known as a maximally correlated states~\cite{Rains1999}.

The second step of the measurement process involves projecting the auxiliary system \(A\) through a PVM. The specific unitary used ensures that the energy eigenstates of \(A\) are correlated with those of \(S\). By performing a projective measurement on \(A\), with the operators defined by the states \(\ket{\psi_n}\),  we indirectly project \(S\) onto its energy eigenspace. Applying the PVM \(\{\mathbbm{1} \otimes \ketbra{\psi_k}\}\) results in the following state
\begin{align}
	\rho_2^{SA} = \sum_k p_k \rho_{2,k}^{SA} = \sum_k p_k (\ketbra{k^S} \otimes \ketbra{\psi_k^A}),
\end{align}
For each outcome \(k\), obtained with probability  \(p_k= \rho_{k,k}\), the state of \(S\), after tracing out the degrees of freedom of \(A\), becomes
\begin{align}
	\rho_{2,k}^S=\ketbra{k}.
\end{align}
Therefore, with this measurement protocol the state of \(S\) is conditioned as if an energy measurement were performed on it, yielding outcome \(k\) and projecting the state accordingly with the appropriate probability, all while performing the measurement solely on \(A\). We leverage results from~\cite{morroneDaemonicErgotropyContinuously2023}, where it is shown that for a rank-one measurement the daemonic ergotropy is always equal to the energy of the unconditional state, regardless of the measurement protocol. Therefore, in the ideal case, we have that 
\begin{align}
    \daem=E(\rho_0^S)=\Tr [ \hat{H}_0 \rho_0^S ] \,.
\end{align}
This measurement scheme constitutes the first block of our protocol, with the only difference being that in the non-ideal case noise is introduced by initializing the auxiliary system \(A\), by considering it in a thermal state.

\subsection{The work cost of quantum measurement}
To balance the work acquired from the measurement, a certain amount of energy must be paid to perform the measurement, and for proper accounting, the work required to erase the acquired information must also be included in the balance. We base our analysis for the work cost of implementing our proposed protocol on the paper~\cite{minagawaUniversalValiditySecond2024} which provides a rigorous treatment of the energy balance under limited assumptions, demonstrating universal bounds to work costs and work gain from generalized measurement protocols. For a detailed derivation of the bounds, we refer the reader to the original paper. The work cost associated with the cost of driving the controller for the measuring protocol in the referred papers is evaluated as
\begin{align}
    W_{\mathrm{in}}^{MK}
&= \beta^{-1}\Delta S^{AMK}_{2,0}
 \nonumber \\
&+\beta^{-1}\!\left[ \mathcal{J}_{GO} + I(A : M \mid K)_{\rho_2} + S^{B_2}_{\mathrm{irr}} \right],
\end{align}
where in their notation the system measured is \(A\) and the controller is \(MK\).
In the analysis performed by Minagawa and colleagues, the controller consists of two subsystem: the memory \(M\) and the classical register \(K\). The former role is to interact with the main system \(A\) so that information can be acquired from the correlations it generates. The result of the measurement is recorded on the classical register, a system in a pure state so that each of the \(k\) outcome of the measurement can be encoded in the \(k\) eigenstates of the system's reference basis. Once the measurement outcome is ``written'' in the classical register, a controlled operation conditioned by its state can be performed on the main system \(A\).
In our framework the auxiliary system \(A\) plays the role of both the memory and the classical register.
This is possible due to our protocol having a less generic scope.
Effectively, while the state of the auxiliary system \(A\) is not pure at all times, due to the measurement of choice being projective, the post measurement state is a pure state conditioned by the measurement outcome.
So the system can fulfill both roles, as long as the post measurement state of \(A\) does not present any ambiguity due to possible degeneracy.
The term \(I(A : M \mid K)_{\rho_2}\) represents the conditional mutual information \(I(A : M \mid K)_{\rho_2}= \sum_k p_k I(A:M)_{\rho_{2,k}}\) between the measured system and the memory. As already discussed by the authors, a sufficient condition for the quantity to vanish is for the post-measurement state to be separable for all outcomes, which is satisfied when all the effects describing the measurement operation are rank-1.
For a PVM this is always satisfied as long as the measured observable is non-degenerate, so no specific assumptions regarding the quantum instruments performing the operation is necessary.
The term \(S^{B_2}_{\mathrm{irr}}\) represents the irreversible entropy production associated with the information erasure step.
We assume that the reset of \(A\) is implemented by an ideal isothermal process, i.e.\ quasi-statically and reversibly, so that \(S^{B_2}_{\mathrm{irr}}=0\).
We stress that a free thermalization of \(A\) with the bath would instead be maximally irreversible, yielding
\(S^{B_2}_{\mathrm{irr}} = S(\rho_2^{A}\|\rho_\beta^{A})\); the reversible
reset recovers the residual free energy of the post-measurement state of the ancilla, and is realized by the same class of isothermal protocols employed in the extraction step.

Under these assumptions, the work cost associated with our protocol can be simplified as
\begin{align}
\label{eq:measurement_cost}
\mathcal{W}_{\mathrm{cost}} &= \beta^{-1} \left( \Delta S_{2,0}^{SA} + \mathcal{J}_{\mathrm{GO}} \right) \\
&=\beta^{-1} \Big[ S(\rho_2^{SA}) - S(\rho_0^S) -S(\rho_\beta^A) \Big] \nonumber \\
& \quad + \beta^{-1} \Big[ S(\rho_0^S) - \sum_k p_k S(\rho_{2,k}^S) \Big] \nonumber \\
&= \beta^{-1} \left[ S(\rho_2^{SA}) - \sum_k p_k S(\rho_{2,k}^S) - S(\rho_\beta^A) \right] \nonumber \\
&= \beta^{-1}\left[ H(\{p_k\}) - S(\rho_\beta^A) \right] = \beta^{-1} \left[ S(\rho_2^A) - S(\rho_\beta^A)
\right],
\nonumber
\end{align}
first we used the additivity of entropy for the factorized state $\rho_0^{SA}=\rho_0^S\otimes\rho_\beta^A$; the two contributions $S(\rho_0^S)$ then cancel.
Then, the rank-one projective measurement produces the classical--quantum state $\rho_2^{SA} = \sum_k p_k\, \rho_{2,k}^S \otimes \ketbra{\psi_k^A}$ with mutually orthogonal states $\{\ket{\psi_k^A}\}$, so its entropy is $S(\rho_2^{SA}) = H(\{p_k\}) + \sum_k p_k S(\rho_{2,k}^S)$, 
where $H(\{p_k\}) = -\sum_k p_k\ln p_k$ is the Shannon entropy of the outcomes.
Finally, since $\rho_2^A = \sum_k p_k\ketbra{\psi_k^A}$, we have $S(\rho_2^A)=H(\{p_k\})$.
Thus, under these assumptions, the overall measurement cost reduces to a Landauer-type cost associated with restoring the ancilla to its initial Gibbs state, while the dependence on the initial state of the system enters only through the distribution of the measurement outcomes. One can also notice how in the ideal limit in which the ancilla is initially pure, it reduces to the standard Landauer cost \(\beta^{-1}H({p_k})\) for an efficient measurement~\cite{sagawaMinimalEnergyCost2009b}.

\section{Work extraction via a non-ideal QND-energy measurement protocol}
\label{s:res1}
We now characterize the protocol we propose for extracting work through a QND-energy measurement.
As represented in Fig.~\ref{f:qnd_model}, we consider a main system \(S\) with free Hamiltonian \(\hat{H}_0=\sum_n \varepsilon_n \ketbra{n}\), starting in a possibly active state \(\rho^S_0=\sum_{n,m} p_{n,m} \ketbra{n}{m}\).
The auxiliary system \(A\), which for simplicity we assume to have the same free Hamiltonian as the system \(S\), is initially thermalized with a thermal bath \(B\) at inverse temperature \(\beta\).
Its state is given by the Gibbs state \(\rho^A_\beta=e^{-\beta \hat{H}_0}/Z\), with \(Z=\Tr [e^{-\beta \hat{H}_0}]\). 
We assume that, at the start, the composite system state is factorized and equal to
\begin{align}
	\rho_0^{SA}= \rho^S_0 \otimes \rho^A_\beta.
\end{align}
Our protocol consists of the following steps:
\begin{enumerate}
	\item \textbf{Unitary evolution:} The system \(S\) and the auxiliary system \(A\) evolve unitarily, as in the ideal case, with the unitary being given by Eq.~(\ref{eq:ch6_unitary}). 
	Under the action of this unitary, the system state now evolves as 
	\begin{align}
		\rho_1^{SA} &= \hat{U}_{SA}(\rho^S_0 \otimes \rho_\beta^A) \hat{U}_{SA}^{\dag} \nonumber \\
		&= \sum_{n,m} \rho_{n,m} 
        \left( |n^S\rangle \langle m^S| \otimes e^{-i \varepsilon_n \hat{A}} \rho_\beta^A e^{i \varepsilon_m \hat{A}^\dag} \right).
	\end{align}
	\item \textbf{Energy measurement:} The energy measurement is performed through the set \(\{ \mathbbm{1}\otimes |\psi_k\rangle \langle \psi_k | \}\), where for simplicity, we chose \(\ket{\psi_k}\) to be eigenstates of of the Hamiltonian. Due to the measurement, the state unravels into a convex combination
	\begin{align}
		\rho_2^{SA}=\sum_k p_k \rho_{2,k}^{SA}\,,
	\end{align}
	with
	\begin{align}
		\rho_{2,k}^{SA} 
        %&=\left( \mathbbm{1}^S\otimes |\psi_k^A\rangle \langle \psi_k^A | \right) \left( \rho_2^{SA} \right) \\
		&= \rho_{2,k}^S\otimes |\psi_k^A\rangle \langle \psi_k^A |.
	\end{align}
	\item \textbf{Work extraction:} Work is extracted from the system  \(S\) by performing the optimal operation conditioned on the measurement outcome.
    Differently from the idealized zero-temperature QND protocol, the conditional post-measurement states \( \rho_{2,k}^S \) generally depend on the initial state \(\rho_0^S\).
    Therefore, at finite auxiliary temperature, knowledge of \(\rho_0^S\) is generally required to determine and implement the optimal outcome-dependent extraction operations.
    \item \textbf{Memory reset:} 
    The auxiliary system \(A\) is reset to the Gibbs
    state by a reversible isothermal process in contact with the bath, returning its state to

	\begin{align}
		\rho_4^{AB}= \rho_\beta^A \otimes \rho_\beta^B
	\end{align}
\end{enumerate}

\subsection{Work extraction with isothermal bath-assisted operations}
We first discuss the work extraction step, and evaluate the work gained from the protocol when the extraction is being performed with isothermal bath-assisted quasi-static operations.
In this case, the passive state associated with all the trajectories is simply
\begin{align}
	\rho_{3,k}^S= \rho^S_\beta,
\end{align}
as when extracting work through isothermal bath-assisted operations the final state reached by the system is the state in thermal equilibrium with the bath.

We can now evaluate the average work extracted from the protocol.
The work extracted from the state conditioned by the outcome \(k\) is given by
\begin{align}
	\work(\rho^S_{2,k})=\free(\rho^S_{2,k})-\free(\rho^S_{3,k})=\free(\rho^S_{2,k})-\free(\rho^S_\beta) \,,
\end{align}
and the average work gained due to the protocol is

\begin{align} \small
	\Delta\enhanwork&= \enhanwork-\work(\rho^S_0) \nonumber \\
    &= \sum_k p_k \work(\rho^S_{2,k}) -\work(\rho^S_0) \nonumber \\
	%&= \sum_k p_k \free(\rho^S_{2,k})-\free(\rho^S_\beta) -\work(\rho^S_0) \nonumber \\
    &= \sum_k p_k \free(\rho^S_{2,k})-\free(\rho^S_\beta) -\work(\rho^S_0)+ \nonumber \\
    &\phantom{=\sum_k p_k \free(\rho^S_{2,k})}+\free (\rho^S_0)- \free (\rho^S_0) \nonumber \\
	%&=\free (\rho^S_0)- \free (\rho^S_0) +\sum_k p_k \free(\rho^S_{2,k})-\nonumber \\
    %&\phantom{=\free (\rho^S_0)- \free (\rho^S_0)}-\free(\rho^S_\beta) -\work(\rho^S_0) \nonumber \\
	&= \sum_k p_k \left(E(\rho_{2,k}^S) -E(\rho_0^S) \right)+ \nonumber \\
    &\phantom{==}+ \beta^{-1} \Big( S(\rho_0^S) - \sum_k p_k S(\rho_{2,k}^S)	 \Big) \nonumber \\
    %&\phantom{\sum_k p_k \left(E(\rho_{2,k}^S)\right) + }+ \beta^{-1} \left( S(\rho_0^S) - \sum_k p_k S(\rho_{2,k}^S)	 \right) \nonumber \\
	&=\beta^{-1}\mathcal{J}_{GO},
\end{align}
where we used
\begin{itemize}
	\item \(\rho^S_1=\sum_k p_k\rho_{2,k}^S\), the unconditional state is obtained by average over the state conditioned by the measurement
	\item \(E(\rho^S_1)=E(\rho_0^S)\), due to the unitary being energy preserving
	\item \(\mathcal{J}_{GO}=S(\rho_0^S) - \sum_k p_k S(\rho_{2,k}^S)\), we substituted the Groenewold-Ozawa information gain.
\end{itemize}
We highlight that the work gain is no longer guaranteed to be positive, as the Groenewold-Ozawa information gain can be negative.
Intuitively, in our protocol \(SA\) correlation needs to be created, and whenever the information acquired from the measurement cannot compensate the cost of correlating the two systems, the total protocol contribution is negative.

Let us define the net daemonic thermal work gain, accounting for the measurement cost, as
\begin{align}
	\Delta\overline{\mathcal{W}}^\beta_{\text{net}} = \Delta \enhanwork - \mathcal{W}_{\text{cost}}\,.
\end{align}
We then obtain the formula
\begin{align}
	\Delta\overline{\mathcal{W}}^\beta_{\text{net}}=- \beta^{-1}  \Delta S_{2,0}^{SA}.
    \label{eq:DeltaWnet}
\end{align}

As discussed by Minagawa and colleagues ~\cite{minagawaUniversalValiditySecond2024}, while in general \(\Delta S_{2,0}^{SA}\) can be negative, a sufficient condition for it to be non-negative is if the measurement applied to the memory is a PVM. As we perform a energy measurement on \(A\), we conclude that it is not possible for the net work gain to be positive.

\subsection{Work extraction with unitary operation}
When extracting with unitary operation, the passive state associated with each conditional state $\rho_{2,k}$ will - in general - depend on the measurement outcome
$\rho_{3,k}^S=\tau_{\rho_{2,k}}^S$.
We can evaluate the change in ergotropy due to the protocol as:
\begin{widetext}
\begin{align}
	\Delta \daem &= \sum_k p_k \erg(\rho^S_{2,k})-\mathcal{E}(\rho^S_0)= \sum_k p_k \left( \mathcal{F}_\beta(\rho_{2,k}^S) - \mathcal{F}_\beta(\tau_{\rho_{2,k}}^S) \right) - \left( \mathcal{F}_\beta(\rho_{0}^S) - \mathcal{F}_\beta(\tau_{\rho_{0}}^S) \right) \nonumber \\
	&= \sum_k p_k \Big[  \left( E(\rho^S_{2,k}) - E(\rho^S_{0}) \right)+\beta^{-1} \left( S(\rho^S_{0})-S(\rho^S_{2,k}) \right) \Big]  + \Big( \mathcal{F}_\beta(\tau_{\rho_{0}}^S) - \sum_k p_k \mathcal{F}_\beta(\tau_{\rho_{2,k}}^S) \Big)  \nonumber \\
	&= E(\rho^S_{1}) - E(\rho^S_{0}) +  \beta^{-1}\mathcal{J}_{GO} + (\mathcal{F}_\beta(\tau_{\rho_{0}}^S)- \sum_k p_k \mathcal{F}_\beta(\tau_{\rho_{2,k}}^S)) \nonumber \\
	&= \beta^{-1}\mathcal{J}_{GO} + \Delta \overline{\mathcal{F}}_{2,0,p} \; , \label{eq:daemonic_gain}
\end{align}  
\end{widetext}
where we defined the quantity $\overline{\mathcal{F}}_{2,0,p} = \mathcal{F}_\beta(\tau_{\rho_{0}}^S) - \sum_k p_k \mathcal{F}_\beta(\tau_{\rho_{2,k}}^S) $, which quantify the average change in the residual free energy of the passive state before and after the protocol. 

It is straightforward to show that the daemonic gain result for our protocol is compatible with the generic bound for work extraction established by Minagawa and colleagues. However, as it might be expected, it does not saturate this bound (in general). As the authors of~\cite{minagawaUniversalValiditySecond2024} discussed, it is generally not possible to saturate the bound with purely unitary operations, and the difference from the upper bound in our case is proportional to 
\begin{align}
	\beta^{-1} I(S|K) = \beta^{-1}\left(S(\sum_k p_k \tau_{\rho_{2,k}}^S) - \sum_k p_k S	(\tau_{\rho_{2,k}}^S)\right).
\end{align} 
The quantity above, as it can be easily checked and as discussed by the authors, goes to zero whenever the relation \(\rho^S_{2,k_1,p}=\rho^S_{2,k_2,p} \; \forall k_1,k_2 \), which is always true in the case of isothermal bath-assisted operations, but in general it is not when extracting with unitary operations.
Noticeably, it becomes true in the ideal zero-temperature scenario, as the passive states for each trajectories simply becomes the system's ground state.

In this case we evaluate the daemonic net gain as
\begin{align}
\Delta\overline{\erg}_{\textrm{net}}= \Delta \daem - \mathcal{W}_{\textrm{cost}}= \Delta \overline{\mathcal{F}}_{2,0,p} -  \beta^{-1}\left[\Delta S^{SA}_{2,0} \right]\,.
\end{align}
Contrarily to the net gain evaluated for bath-assisted operations, the net daemonic gain can be either positive or negative.

\section{Measurement-enhanced work extraction from a qubit quantum system}
\label{s:res2}
In this section, we apply
our work extraction protocol to the paradigmatic example of a qubit-qubit system.
Specifically, we consider a system \(S\), modeled as a two-level system with Hamiltonian 
\begin{align}
    \hat{H}_0^S = \omega_0(\hat{\sigma}_z^{S} + \mathbbm{1})/2 \,,
    \label{eq:qubitHamiltonian}
\end{align}
initialized in an arbitrary state \(\rho_0^S= (\mathbbm{1}+ \vec{v} \cdot \vec{\sigma}^S )/2 \), which is completely determined by the Bloch vector \( \vec{v} = (v_x, v_y, v_z) \) whose components correspond to the average values of the system Pauli matrices $\vec{\sigma}^S = (\sigma_x^S,\sigma_y^S,\sigma_z^S)$. 
Throughout this section, all energies and the bath temperature are expressed in units of $\omega_0$ (effectively setting $\omega_0 = 1$).
The auxiliary system \(A\) also consists of a qubit, with Hamiltonian of the same form of \eqref{eq:qubitHamiltonian}, which is initially prepared in a thermal state at inverse temperature \(\beta\). The QND operator \(\hat{A}\), satisfying the relation in Eq.~\eqref{eq:ch6_unitary_cond}, is chosen as \(\hat{A} = \pi \hat{\sigma}_x^A/2\) (because of this choice, the measurement basis \(\ket{\psi_n} \) corresponds to the $\hat{\sigma}_z^A$ (energy) eigenbasis of the auxiliary system \(A\) ).
It is possible to obtain analytical formulas for all the figures of merit we introduced in the previous sections, which will be in general functions of the Bloch vector $\vec{v}$ components, and of the bath inverse temperature $\beta$. 

To make the obtained formulas more readable and instructive, we introduce here few additional parameters and functions:\\ 
i) the modulus of the initial state Bloch vector:
\begin{align}
    v = \sqrt{v_x^2+v_y^2+v_z^2} \,,
\end{align}
which is bounded as $0\leq v \leq 1$.\\
ii) the probability of the auxiliary system being in the excited state 
\begin{align}
\label{eq:pth_def}
p_{\mathrm{th}}(\beta) = \frac{1}{1+e^\beta} \,,
\end{align}
which can take values $0\leq p_{\mathrm{th}}(\beta) \leq 1/2$.\\ 
iii) the shifted binary entropy function
\begin{align}
\phi(x)= \left(\frac{1+x}{2}\right) \log(1+x) + \left(\frac{1-x}{2}\right) \log(1-x) \,.
\end{align}
Furthermore, without loss of generality, we will often implicitly assume $v_y = 0$. Owing to the symmetry of the problem, coherences in the energy eigenbasis associated with the $v_x$ and $v_y$ components of the Bloch vector contribute identically to all the quantities considered here.
We can therefore restrict our analysis to $v_x$ alone as the parameter characterizing the input state coherences.
\subsection{Measurement cost}
 \begin{figure}[b!]
		\includegraphics[width=0.48\textwidth]{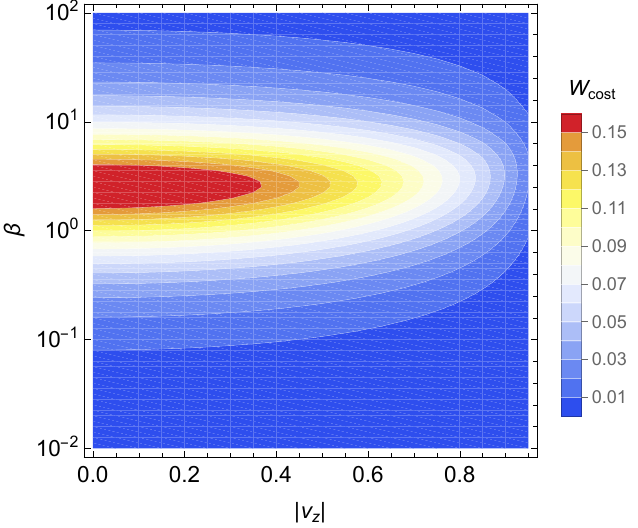}  
        \caption{Measurement cost $\mathcal{W}_{\sf cost}$ for the non-ideal QND measurement on a qubit initial state $\rho_0^S$ as a function of the absolute value of its Bloch component $|v_z|$ and of the inverse temperature $\beta$ caractherizing the thermal state of the auxiliary quantum system.}
        \label{f:Wcost_vs_vz_beta}
 \end{figure}
As described in Eq.~\eqref{eq:measurement_cost}, the measurement cost $\mathcal{W}_{\textrm{cost}}$ is obtained as a sum of the global entropy variation $\Delta S_{2,0}^{SA}$ and the Groenewold-Ozawa mutual information $\mathcal{J}_{GO}$.

As regards the entropy variation, one obtains
 \begin{align}
   \Delta S_{2,0}^{SA}&= \phi(v) - \phi(v_z) \,.
   \label{eq:entropy_variation}
   % \frac{1}{2} \log \left(\frac{v^2-1}{v_z^2-1}\right)-v_z \tanh ^{-1}\left(v_z\right) %\nonumber \\
   %&\,\,\,\,\,\,\, + v \tanh ^{-1}(v).
 \end{align}
Since $\phi(x)$ is an increasing function of its argument and $v_z \leq v$, we find that $\Delta S_{2,0}^{SA}$ is, as expected, always non-negative.
Moreover, we observe that it does not depend on the temperature $\beta$ of the auxiliary system and it goes to zero whenever the system initial state has no coherences in the energy eigenbasis (such that $v=v_z$). More in general, $\Delta S_{2,0}^{SA}$ monotonically increases with the coherence parameter $|v_x|$, being maximized for initial eigenstates of $\hat\sigma_x$, that is for $v_x=\pm 1$. 

Also the Groenewold-Ozawa mutual information $\mathcal{J}_{GO}$, which quantifies the amount of information that one obtains from the measurement, can be derived analytically, obtaining
 \begin{align}
   \mathcal{J}_{GO}&= \phi(k_\beta)+\phi(v_z) - \phi(k_\beta v_z) - \phi(v) \,,
   \label{eq:IGO}
 \end{align}
where we have introduced the additional parameter
\begin{align}
k_\beta = 1-2 \,p_{\mathrm{th}}(\beta) = \tanh (\beta/2) \,,
\end{align}
 which is bounded as $0\leq k_\beta \leq 1$.
From the formula above, it is then apparent that $\mathcal{J}_{GO}$ may take negative values in the presence of coherences ($v_z<v$) and in general for high temperaturs (low $\beta$, and thus small $k_\beta$). On the other hand, $\mathcal{J}_{GO}$ is always non-negative for input states diagonal in the energy eigenbasis.

By summing these two quantities one obtains the measurement cost, whose formula simply reads
 \begin{align}
   \mathcal{W}_{\textrm{cost}} &= \beta^{-1}\left( \phi(k_\beta) - \phi(k_\beta v_z) \right)  \,.
   \label{eq:measurement_cost_qubit}
 \end{align} 
which indeed corresponds to~Eq.~\eqref{eq:measurement_cost}.

We observe that $\mathcal{W}_{\textrm{cost}}$ depends only on the bath temperature $\beta$ and on the absolute value of the z-component of the Bloch vector $|v_z|$ (we remind that $\phi(x)=\phi(-x)$). The behaviour as a function of these parameters can be observed in the contour plot in Fig.~\ref{f:Wcost_vs_vz_beta}.  In particular, the measurement cost is monotonically decreasing with $|v_z|$, being equal to zero for $v_z=\pm 1$ and maximized for $v_z=0$, while it is not a monotonic function of $\beta$. 
In particular, because of the $\beta^{-1}$ factor, the measurement cost goes to zero for zero temperature ($\beta\to\infty$), in accordance with the Landauer cost, which is indeed zero if one has at disposal a zero-temperature bath. Furthermore, the same behaviour is observed also for infinite temperature ($\beta \to 0$). In fact one can easily prove that
\begin{align}
\lim_{\beta \to 0} \mathcal{J}_{GO} &=  \phi(v_z) - \phi(v) = - \Delta S_{2,0}^{SA} \,.
\end{align}
Following the observation made above on $ \Delta S_{2,0}^{SA}$ reported in Eq.~\eqref{eq:entropy_variation}, we can thus conclude that for large temperatures and in the presence of coherences, the Groenewold-Ozawa mutual information takes negative values, attaining in the infinite temperature limit exactly the opposite the entropy variation, and thus leading to a zero measurement cost.

For a given inverse temperature $\beta$, one can directly evaluate the maximum measurement cost over all possible initial states by fixing $v_z=0$, obtaining
 
 \begin{align}
     \mathcal{W}_{\textrm{cost,max}}(\beta) = \frac{\phi(k_\beta)}{\beta} %\nonumber \\
     = \frac{1}{2}\left[
     \tanh (\frac{\beta}{2}) - \frac{2}\beta \log \,\cosh(\frac{\beta}{2})
     \right],
     % \frac{1}{2}- \frac{1}{1+e^\beta} + \frac{\log{2} -\log(1+\cosh\beta)}{2 \beta}\,,
 \end{align}
 which is in turn maximized for $\beta \approx 2.55$, such that $\mathcal{W}_{\textrm{cost,max}}\approx 0.17$.
 Finally, as clearly depicted in Fig.~\ref{f:Wcost_vs_vz_beta}, it is worth noting that this specific inverse temperature $\beta \approx 2.55$ aprroximately maximizes the measurement cost for every initial state, i.e. independently of the parameter $v_z$.
\subsection{Daemonic extracted work and daemonic gain}
\begin{figure*}
	\centering
    \includegraphics{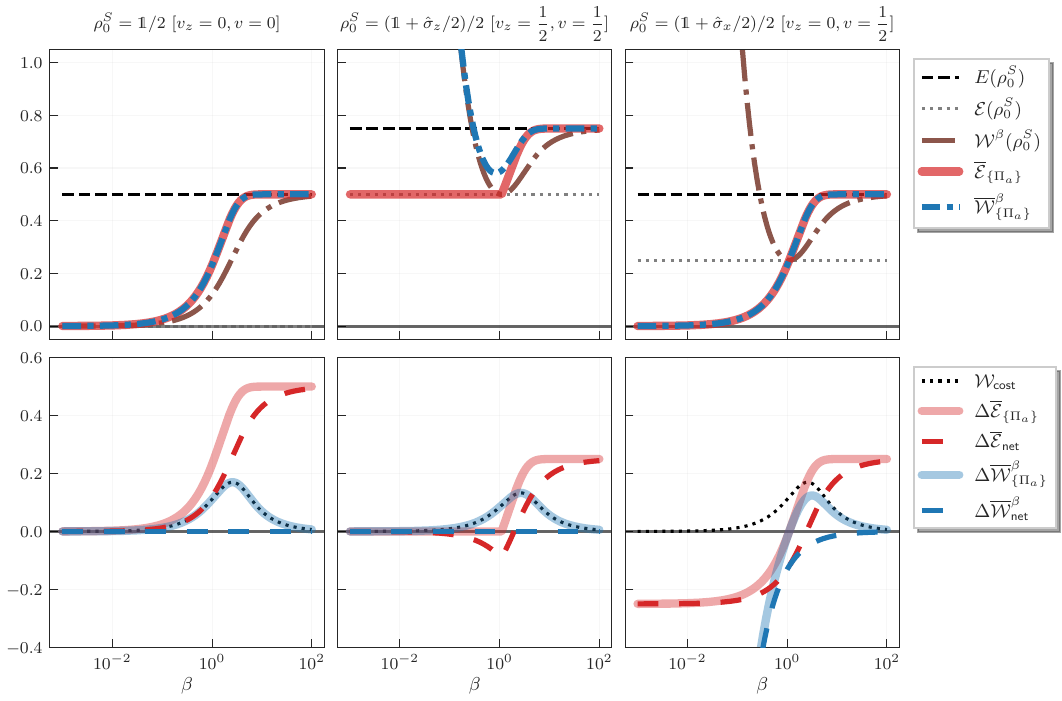}
    \caption{We plot the different figures of merit we have presented for the qubit QND work extraction protocol, considering three different possible initial states: left panels - maximally mixed state: $\rho_0^S = \mathbbm{1}/2$; middle panels - active initial state diagonal in the energy eigenbasis $\rho_0^S = (\mathbbm{1} + \hat{\sigma}_z/2)/2$; right panels - active initial state with coherences in the energy eigenbasis: $\rho_0^S = (\mathbbm{1} + \hat{\sigma}_x/2)/2$. In the top panels we focus on the absolute values of the work extractable via the QND-daemonic protocol, while in the bottom panel we focus on the enhancement obtainable via these protocols, with and without considering the measurement cost in the energy balance.}
    \label{f:DaemonicQubit}
\end{figure*}
We will here address specifically the absolute values of the daemonic work extracted via thermal and unitary operations, along with the corresponding daemonic enhancement, respect to extraction protocols not exploiting a QND measurement.
Besides deriving the corresponding analytical formulas, in Fig.~\ref{f:DaemonicQubit} we show the behaviour of such quantities for three different input states of the system: the left panels correspond to the maximally mixed state, $\rho_0^S = \mathbbm{1}/2$; the middle panels refer to an active state diagonal in the energy eigenbasis, $\rho_0^S = (\mathbbm{1} + \hat{\sigma}_z/2)/2$; the right panels correspond to an active state with coherences in the energy eigenbasis, $\rho_0^S = (\mathbbm{1} + \hat{\sigma}_x/2)/2$.\\

As regards the work extractable via bath-assisted operations, one obtains the following analytical formulas
\begin{align}
\work (\rho_0^S) &= \frac{v_z}{2} + \frac{\phi(v) + \log(\cosh(\beta/2))}{\beta} \,, \\
\enhanwork &= \frac{v_z + 1}{2} - p_{\mathrm{th}}(\beta) + \frac{\phi(v_z) - \phi(k_\beta  v_z)}{\beta} \,.
\end{align}
As one can see in the top panels of Fig.~\ref{f:DaemonicQubit}, both functions are in general not monotonic in $\beta$; they both tend to the asymptotic value $E(\rho_0^S)=(v_z + 1)/2$ in the limit of zero temperature ($\beta \to \infty$), while their behaviour at high temperatures depends on the specific input states: $ \work (\rho_0^S)$ goes to zero for $\beta\to 0$ only for a maximally mixed input state (that is for $v=0$, see top-left panel of Fig.~\ref{f:DaemonicQubit}), otherwise it diverges and presents a minimum at a temperature $\beta^*$ such that the input state and the auxiliary system thermal state have the same von-Neumann entropy $S(\rho_0^S)=S(\rho_{\beta^*}^A)$, corresponding to the condition $p_{\mathrm{th}}(\beta^*) = (1-v)/2$. The daemonic thermal work $\enhanwork$ diverges for high temperatures whenever $v_z \neq 0$, as one can notice in the top-center and top-right panels of Fig.~\ref{f:DaemonicQubit}, and it goes to zero for $v_z=0$, as shown in the top-middle panel of Fig.~\ref{f:DaemonicQubit}. 

We remind ourselves here that the daemonic thermal gains (excluding and including the measurement cost) can be written in terms of the quantities discussed above, that is
\begin{align}
    \Delta \overline{\mathcal{W}}^\beta_{\{\Pi_a^A\}} &= \beta^{-1} \mathcal{J}_{GO}  \\
    &=\beta^{-1} \left( \phi(k_\beta)+\phi(v_z) - \phi(k_\beta v_z) - \phi(v) \right) \,,  \nonumber
    \\
\Delta\overline{\mathcal{W}}^\beta_{\text{net}} &= - \beta^{-1} \Delta S_{2,0}^{SA} 
\nonumber \\
    &= - \beta^{-1} \left( \phi(v) - \phi(v_z) \right) \,.
\end{align}
As a consequence, the observations made before on $\mathcal{J}_{GO}$ and $\Delta S_{2,0}^{SA}$ can be translated here, with the temperature acting simply as a scale parameter.
In summary, the thermal daemonic gain $\Delta \overline{\mathcal{W}}^\beta_{\{\Pi_a^A\}}$ is nonnegative for initial state diagonal in the energy eigenbasis, being in these cases in fact equal to measurement cost $\Delta \overline{\mathcal{W}}^\beta_{\{\Pi_a^A\}}=\mathcal{W}_{\sf cost}$ (see bottom-left and bottom-middle panels of Fig.~\ref{f:DaemonicQubit}).
On the other hand, as one can observe from the bottom-right panel of Fig.~\ref{f:DaemonicQubit}, it may take negative values for initial states with coherences and for high enough temperature.
If one also includes the cost of the measurement, then one obtains that $\Delta\overline{\mathcal{W}}^\beta_{\text{net}}$ is, as expected, non positive, equal to zero for states with no coherences, while it monotonically decreases with the coherence parameter $|v_x|$, being minimized for initial states corresponding to the eigenstates of $\hat\sigma_x$, that is for $v_x=\pm 1$.\\

In the case of unitary work extraction operations, one obtains the following simple analytical results for the no-measurement and daemonic ergotropy:
\begin{align}
    \mathcal{E}(\rho_0^{S}) &= \frac{v + v_z}{2} \,,\nonumber \\
    \daem &= \frac{ \left|v_z + 1 -2 p_{\mathrm{th}}(\beta) \right| + \left|v_z-1 + 2 p_{\mathrm{th}}(\beta)\right| + 2 v_z}{4} \,,\nonumber 
    %&=\frac{1}{4} \left(\frac{\left| e^{\beta } \left(v_z-1\right)+v_z+1\right| +\left| v_z+e^{\beta }
   %\left(v_z+1\right)-1\right| }{e^{\beta }+1} \right. \nonumber \\
   %& \,\,\,\,\,\,\,\, +2 v_z \Bigg).
\end{align}
where $p_{\mathrm{th}}(\beta)$ is the probability of the auxiliary system being in the excited state, as defined in Eq.~\eqref{eq:pth_def}.
From the formula above one can immediately prove that a new threshold temperature $\beta_z^*$ exists, defined via the relation  $p_{\mathrm{th}}(\beta_z^*)= (1 - |v_z|)/2$, such that
\begin{align}
    \daem =
        \begin{cases}
        \max(0,v_z) & \text{if } \beta<\beta_z^* \,, \\
        \frac{v_z+1}{2} - p_{\mathrm{th}}(\beta) & \text{if } \beta > \beta_z^* \,,
        \end{cases}
\end{align}   
showing how in the limit of zero temperature ($\beta \to \infty$) also the daemonic ergotropy goes to the limit value $E(\rho_0^S)=(v_z+1)/2$, as clearly shown in all the top panels of Fig.~\ref{f:DaemonicQubit}.
Furthermore one can also demonstrate that for initial states such that $v_z=0$, then 
\begin{align}
\overline{\mathcal{E}}_{\{ \Pi_a^A \}} &= \overline{\mathcal{W}}^\beta_{\{\Pi_a^A\}} = \frac{1}{2} - p_{\mathrm{th}}(\beta) \,\,\,\,\,\,\text{(if $v_z=0$)} \,,
\end{align}
that is for states having equal population of ground and excited states, unitary operations are as efficient as bath-assisted operations in the QND-daemonic protocol (see top-left and top-right panels of Fig.~\ref{f:DaemonicQubit}). 

More in general, we observe that, while the ergotropy of the initial state $\mathcal{E}(\rho_0^{S})$ depends both on the energy population $v_z$ and on the coherences through the Bloch vector $v$, the daemonic erogtropy depends only on $v_z$ and on the temperature $\beta$. Effectively, the measurement protocol eliminate any coherence in the energy eigenbasis and, as they do not influence the measurement outcome distribution, they have no influence over the average extracted work. 
It is then trivial to derive the daemonic ergotropy gain, which reads
\begin{align} 
    \Delta\overline{\mathcal{E}}_{\{ \Pi_a^A \}} &= 
    \frac{ \left|v_z + 1 -2 p_{\mathrm{th}}(\beta) \right| + \left|v_z-1 + 2 p_{\mathrm{th}}(\beta)\right| - 2 v}{4} \nonumber \\
    %\frac{1}{4} \left(\frac{\left| e^{\beta } \left(v_z-1\right)+v_z+1\right| +\left| v_z+e^{\beta }
   %\left(v_z+1\right)-1\right| }{e^{\beta }+1} \right. \nonumber \\
   %& \,\,\,\,\,\,\,\, -2 v \Bigg)\,,
    &= \frac{\max(k_\beta, |v_z|) - v}{2}  \nonumber \\
    &=\begin{cases}
        \frac{|v_z|-v}{2} & \text{if } \beta<\beta_z^* \\
        \frac{1-v}{2}-p_{\mathrm{th}}(\beta)  & \text{if } \beta > \beta_z^*
    \end{cases}
    \label{s:daemgain}
\end{align}
and thus includes a clear dependence on the Bloch vector components $v_x$ and $v_y$, which negatively affect its value. In principle also this quantity may take negative values: from the formula above it is clear that $\Delta\overline{\mathcal{E}}_{\{ \Pi_a^A \}} > 0$ if and only if $p_{\mathrm{th}}(\beta) < (1-v)/2$, and thus implying the existence of a threshold temperature $\beta=\beta^*$ corresponding to the 
equivalence $S(\rho_0^S)=S(\rho_{\beta^*}^A)$ between the von-Neumann entropy of the input state and the one of the auxiliary thermal bath state, and thus to the minimum of $\work (\rho_0^S)$. One should notice that in general $\beta^* \geq \beta_z^*$, and we refer to the middle and right panels of Fig.~\ref{f:DaemonicQubit} for examples of such behaviour. Remarkably one also observes that for $\beta<\beta^*$, then $\Delta\overline{\mathcal{E}}_{\{ \Pi_a^A \}}=0$ if the initial state is diagonal in the energy eigenbasis ($v=v_z$), while it takes negative values in the presence of coherences ($|v_z|<v$). Notice that, trivially, for input states $\rho_0^S$ diagonal in the energy eigenbasis one has $v=|v_z|$ and thus $\beta^*=\beta_z^*$.

The analytical formula for the net-daemonic-ergotropy gain $\Delta\mathcal{E}_{\sf net}$ can be readily obtained from the ones reported above, yielding
\begin{align}
    \Delta\mathcal{E}_{\sf net}&= \frac{\max(k_\beta, |v_z|) - v}{2} - \beta^{-1}\left(\phi(k_\beta) - \phi(k_\beta v_z)\right) \,.
\end{align}
A few crucial observations can be made: since in general $\Delta\mathcal{E}_{\sf net}\leq \Delta\overline{\mathcal{E}}_{\{ \Pi_a^A \}}$, then also the net-daemonic-ergotropy gain may take both negative and positive values.
The corresponding threshold temperature $\beta_{\sf net}^*$ such that $\Delta\mathcal{E}_{\sf net}=0$ cannot be obtained analytically. However it clearly satisfies $\beta_{\sf net}^* \geq \beta^* \geq \beta_z^*$, and can be evaluated numerically for specific input states $\rho_0^S$. Furthermore, one can also prove that the previously introduced threshold temperature $\beta_z^*$ plays an interesting role regarding the behaviour of the net-daemonic gain as a function of the temperature: for $\beta< \beta_z^*$, $\Delta\mathcal{E}_{\sf net}$ is monotonically decreasing with $\beta$, while for $\beta>\beta_z^*$ it becomes monotonically increasing (cf. fthe bottom-middle panel of Fig.~\ref{f:DaemonicQubit}).

\section{Conclusions \& outlooks }
\label{s:conclusion}
We studied measurement-assisted work extraction based on a QND energy measurement in the presence of a finite-temperature bath, explicitly accounting for the energetic cost of performing the measurement and resetting the measurement apparatus. In our model, the finite bath temperature not only degrades the quality of the measurement, but also introduces a non-zero energetic cost associated with its implementation.

Our main objective was to compare two operational settings: one in which work extraction is restricted to unitary operations, and one in which access to a thermal bath is also allowed. The central question was whether the information gained through the measurement can still provide a net thermodynamic advantage once its energetic cost is taken into account.

Our most important result concerns the corresponding net gains. We proved that, when access to a thermal bath is allowed, the daemonic net gain is always non-positive, independently of the bath temperature and of the input state. Therefore, once the energetic cost of performing and resetting the measurement is properly included, no thermodynamic advantage can be obtained from the QND-assisted protocol in this operational setting. This result provides a rigorous limitation on the usefulness of measurement-assisted extraction when access to a thermal bath is already available.

The situation is markedly different when work extraction is restricted to unitary operations. In this case, the measurement can still provide a genuine operational advantage, and positive values of the net gain become possible. More generally, we showed that measurement-assisted unitary extraction can outperform the corresponding no-measurement benchmark and, in some parameter regimes, can even extract more work than thermal protocols that do not employ measurements. Our analysis identified the origin of this behaviour in an additional contribution associated with the residual free energy of the conditional passive states generated by the measurement process.

These general findings were illustrated through a detailed study of a qubit system measured by a qubit auxiliary system. Besides confirming the existence of positive net-gain regimes for unitary extraction, the qubit analysis revealed interesting situations in which, for states diagonal in the energy basis, the measurement-assisted unitary protocol can achieve the same extracted work as the corresponding measurement-assisted thermal protocol. While this feature may rely on specific properties of two-level systems and does not necessarily generalize to higher-dimensional systems, it provides useful insight into the interplay between information acquisition, feedback, and work extraction. Moreover, we identified several temperature thresholds with distinct operational meanings, highlighting the parameter regimes in which daemonic work-extraction protocols can provide an advantage over their measurement-free counterparts.

Our model is also closely connected to realistic experimental platforms. In particular, QND measurements of superconducting qubits can be implemented through dispersive interactions with cavity modes coupled to thermal environments, where increasing thermal occupation is known to reduce the measurement efficiency~\cite{Cernotik2015}. The framework developed here thus provides a natural starting point for experimental investigations of the energetic balance of measurement-assisted work extraction.

Several directions for future research remain open.
One possibility is to extend the present framework to sequential or continuous-measurement scenarios, for example through collisional models~\cite{ciccarelloQuantumCollisionModels2022a,landiInformationalSteadyStates2022,elyasiExperimentalSimulationDaemonic2024} or stochastic-master-equation~\cite{albarelliPedagogicalIntroductionContinuously2024} descriptions of weak measurements.
Another interesting direction concerns the idealization of the thermal bath: since the bound~\eqref{eq:thermalwork} is tight only for an infinitely large bath, it would be interesting to investigate the role of baths with a finite heat capacity in daemonic work-extraction protocols.
Finally, one may consider other classes of operations for work extraction, in particular those formulated within a resource-theoretic framework.
Thermal operations and the experimentally motivated subclass of Elementary Thermal Operations (ETOs)~\cite{lostaglioElementaryThermalOperations2018}, impose energy-conservation and covariance constraints that are absent in the isothermal setting considered here; addressing measurement-assisted extraction in that framework would require both an explicit work-storage system and a separate treatment of coherences.

% %
\section*{Acknowledgments}
MGG acknowledges useful discussions with Francesco Buscemi.
MP acknowledges support from the Royal Society Wolfson Fellowship (RSWF/R3/183013), the Department for the Economy of Northern Ireland under the US-Ireland R\&D Partnership Programme, and the EU Horizon Europe EIC Pathfinder project QuCoM (GA no. 10032223).
DM and MP are supported by the PNRR PE Italian NQSTI (PE0000023).

\bibliography{bibliography}

\appendix
\onecolumngrid

\end{document}